\documentclass{emulateapj}

\shorttitle{Phases of Decay Rate Variations}
\shortauthors{Sturrock et al.}

\begin{document}


\title{Concerning the Phases of Annual Variations of Nuclear Decay Rates}


\author{P.A. STURROCK\altaffilmark{1}, J.B. BUNCHER\altaffilmark{2,3}, E. FISCHBACH\altaffilmark{2}, D. JAVORSEK II\altaffilmark{4},
 J.H. JENKINS\altaffilmark{5,2}, AND J.J. MATTES\altaffilmark{2}}
%
%
%


\altaffiltext{1}{Center for Space Science and Astrophysics, Stanford University,
Stanford, CA 94305; sturrock@stanford.edu}
\altaffiltext{2}{Department of Physics, Purdue University, West Lafayette, IN 47907}
\altaffiltext{3}{Department of Physics, Wittenberg University, Springfield, Ohio 45501 USA}
\altaffiltext{4}{416th Flight Test Squadron, 412th Test Wing, Edwards AFB, CA 93524, USA}
\altaffiltext{5}{School of Nuclear Engineering, Purdue University, West Lafayette, IN 47907}


\begin{abstract}
Recent analyses of datasets acquired at the Brookhaven National Laboratory and at the Physikalisch-Technische Bundesanstalt both show evidence of pronounced annual variations, suggestive of a solar influence. However, the phases of decay-rate maxima do not correspond precisely to the phase of minimum Sun-Earth distance, as might then be expected. We here examine the hypothesis that decay rates are influenced by an unknown solar radiation, but that the intensity of the radiation is influenced not only by the variation in Sun-Earth distance, but also by a possible North-South asymmetry in the solar emission mechanism. We find that this can lead to phases of decay-rate maxima in the range 0 to 0.183 or 0.683 to 1 (September 6 to March 8) but that, according to this hypothesis, phases in the range 0.183 to 0.683 (March 8 to September 6) are ``forbidden.'' We find that phases of the three datasets here analyzed fall in the allowed range. 
\end{abstract}


\keywords{Methods: data analysis, Sun: particle emission}



\section{Introduction}

\citet{jen09} and \citet{fis09} have drawn attention to the fact that decay-rate measurements made at the Brookhaven National Laboratory (BNL; \cite{alb86}) and at the Physikalisch-Technische Bundesanstalt (PTB; \cite{sie98}) show clear evidence of periodic annual modulations. According to their analyses, the decay rates appear to peak in winter, suggesting that some kind of radiation from the Sun is influencing the decay rates. However, the dates of maximum rate differ significantly from January 3, which is the date of minimum Sun-Earth distance, which argues against a simple model in which decay rates are solely influenced by an unidentified but isotropic radiation from the Sun. The purpose of this article is to propose an interpretation of the phases of the annual oscillations.

The suggestion that nuclear decay rates may be variable has of course been questioned by a number of other groups:
\begin{enumerate}
\item{}\citet{sem09} and others suggested that these fluctuations have their origin in environmental influences on the detector systems, or other systematic effects. However, \citet{jen10} have shown that the results of their BNL and PTB analyses cannot be explained by variations of temperature, pressure, humidity, etc.
\item{}\citet{nor09}, re-examined data from several nuclear-decay experiments, but found no evidence for a correlation with Sun-Earth distance. However, we have re-analyzed data for $^{22}$Na and $^{44}$Ti decay rates, which Norman and his collaborators generously provided to our collaboration. Norman and his colleagues compared these data with a sinusoidal modulation of fixed amplitude 0.15\%, and fixed phase of year (0.01), to conform with the results found by \citet{jen09} in their analyses of BNL and PTB data. However, we have carried out a power spectrum analysis of the Norman data, which searches for evidence of periodicity and yields estimates of the amplitude and phase of any oscillation \citep{oke11}. Our analysis yields evidence of an annual modulation, with significance estimated at 1\%.  We estimate the amplitude to be 0.034\% [smaller than that considered by Norman et al. by a factor of 5], and the phase to be 0.06 $\pm$ 0.05, which is compatible with the value [0.01] assumed by Norman et al. The fact that the amplitude inferred from the data of Norman, et al. is smaller than what is observed in the BNL or PTB data is compatible with the standard weak interaction theory, as noted by \citet{jen10}.
\item{}\citet{coo09} analyzed data from the power output of the radioisotope thermoelectric generators (RTGs) aboard the Cassini spacecraft, but found no significant deviations from exponential decay. However, \citet{jen10} have shown that there is in fact no conflict between Cooper's results and their results. This is due in part to the fact that the Cassini RTGs derive their power from the alpha decay of $^{238}$Pu, whereas the periodic effects observed in various data sets always involve beta decays.
\end{enumerate}

We have recently presented evidence that both BNL and PTB measurements also exhibit a periodicity suggestive of the influence of solar rotation--possibly of the solar core \citep{stu10a,stu10b}. Such a periodicity would indicate that the radiation is anisotropic, indicating specifically a longitudinal (East-West) asymmetry. If the radiation has a longitudinal asymmetry, it is reasonable to suspect that it may also have a latitudinal (North-South) asymmetry. 

The goal and scope of this article are limited to investigating the consequences of the following two hypotheses:
\begin{enumerate}
\item[(a)]\textit{Some nuclear decay rates are influenced by an unknown solar radiation;} and
\item[(b)]\textit{The flux of this radiation, as it might be measured at Earth, is influenced by two factors: the variation in the Sun-Earth distance due to the eccentricity of the Earth's orbit, and a North-South asymmetry in the solar radiation.}
\end{enumerate}

In Section \ref{sec:Model}, we analyze the properties of a model based on these two hypotheses. We find that, according to this model, the maximum decay rate should occur in the range approximately September 6 to March 8. In Section \ref{sec:data}, we compare this model with data from the BNL and PTB experiments, and find that the phases of maximum variability appear to be compatible with this model. We discuss these results in Section \ref{sec:disc}.

\section{Model}\label{sec:Model}


It is convenient to measure phase of the year so as to run from 0 to 1. We denote by $\phi_o$ the phase of the minimum Sun-Earth distance, and by $\phi_A$ a phase related to the solar axis of rotation. Since the Sun-Earth distance is a minimum on January 3, $\phi_o=0.008$. 

The Earth has maximum exposure to the southern solar hemisphere on March 6 and maximum exposure to the northern hemisphere on September 8, corresponding to phases 0.178 and 0.687, respectively. The fact that the difference is not exactly 0.5 is due to the eccentricity of the Earth's orbit. Approximating the elliptical orbit by a circular orbit for the purpose of representing the North-South asymmetry, we adopt $\phi_A=0.183$, which gives maximum exposure to the southern hemisphere at phase 0.183 (March 8) and maximum exposure to the northern hemisphere at phase 0.683 (September 6).

We also point out that we are implicitly invoking a third hypothesis:
\begin{enumerate}
\item[(c)]\textit{Whatever region of the Sun is responsible for a solar influence on decay rates has an axis of rotation that is indistinguishable from that determined by observation of the solar photosphere.}
\end{enumerate}
The eccentricity of the Earth's orbit leads to a variation in the flux of the hypothetical radiation influencing decay rates, normalized to mean value unity, given by

\begin{equation}
F_o=1+C\cos\left[2\pi\left(\phi-\phi_o\right)\right],
\label{eq1}
\end{equation}  
  		
\noindent{}where $C=0.0334$. The variation of the flux due to a North-South asymmetry is given by

\begin{equation}
F_A=1-AC\cos\left[2\pi\left(\phi-\phi_A\right)\right],
\label{eq2}
\end{equation}

where the ``asymmetry coefficient'' $A$ is chosen so that:
\begin{enumerate}
\item[(a)]If $|A|=1$, the amplitude of the variation due to North-South asymmetry is equal to the amplitude of the variation due to the eccentricity of the Earth's orbit, and
\item[(b)]If $A>0$, radiation from the northern hemisphere exceeds that from the southern hemisphere (and vice versa). Since $C$ is small (and if $A$ is not large), the total variation in the flux (retaining the normalization to mean value unit) is given by
\end{enumerate}
\begin{eqnarray}
F=1+C
\left\lbrace\cos\left[2\pi\left(\phi-\phi_o\right)\right]-A\cos\left[2\pi\left(\phi-\phi_A\right)\right]\right\rbrace,
\label{eq3}
\end{eqnarray}
If the average flux is responsible for an increase in the decay rate by the small ``coupling coefficient'' $\Gamma$, then the annual modulation of the decay rate $R$, normalized to mean value unity, is given by
\begin{eqnarray}
R=1+\Gamma{}C
\left\lbrace\cos\left[2\pi\left(\phi-\phi_o\right)\right]-A\cos\left[2\pi\left(\phi-\phi_A\right)\right]\right\rbrace,
\label{eq4}
\end{eqnarray}

\noindent{}[We should note, however, that it is conceivable that the unknown radiation may act on some isotopes to reduce the decay rate. In this case the coupling coefficient $\Gamma$ would be negative.]

We find from inspection of this formula that, since $|\phi_A-\phi_o|<0.5$, the maximum value of $R$ is to be found only in the range $\phi_A - 0.5 ~\rm{to}~\phi_A$ or, equivalently, in the two ranges $0 ~\rm{to}~\phi_A$ and $\phi_A + 0.5 ~\rm{to}~1$, i.e. 0 to 0.183 and 0.683 to 1. The range $\phi_A ~\rm{to}~\phi_A + 0.5$, i.e. 0.183 to 0.683, is ``forbidden.'' [For the case that the solar radiation acts to suppress the decay rate, these ranges would be reversed.] One may understand this result by noting that if we sum two sine waves, with phases $\phi_1 ~\rm{and}~ \phi_2$, the peak will be found in the range $\phi_1~ \rm{to}~ \phi_2$ if $|\phi_1-\phi_2|<0.5$, and outside that range if $|\phi_1-\phi_2|>0.5$.

Expressed in the form in which it would be measured experimentally, this formula
becomes

\begin{equation}
R=1+K\cos\left[2\pi\left(\phi-\kappa\right)\right],
\label{eq5}
\end{equation}

\noindent{}in which $K$ is the amplitude of the variation, and $\kappa$ is the phase of the peak. By comparing Eqs. \ref{eq4} and \ref{eq5} and separating out coefficients of $\cos\left(2\pi\phi\right)$ and $\sin\left(2\pi\phi\right)$, we find that

\begin{eqnarray}
\nonumber{}K\cos\left(2\pi{}\kappa{}\right)&=&\Gamma{}C\left[\cos\left(2\pi{}\phi_o{}\right)-A\cos\left(2\pi{}\phi_A{}\right)\right]\\
K\cos\left(2\pi{}\kappa{}\right)&=&\Gamma{}C\left[\sin\left(2\pi{}\phi_o{}\right)-A\sin\left(2\pi{}\phi_A{}\right)\right]
\label{eq6}
\end{eqnarray}
	
\noindent{}from which we find that the asymmetry coefficient $A$ is related to $\kappa$ by

\begin{equation}
A=\frac{\sin\left[2\pi{}\left(\kappa{}-\phi_o\right)\right]}{\sin\left[2\pi{}\left(\kappa{}-\phi_A\right)\right]}
\label{eq7}
\end{equation}

\begin{figure}[ht]
\plotone{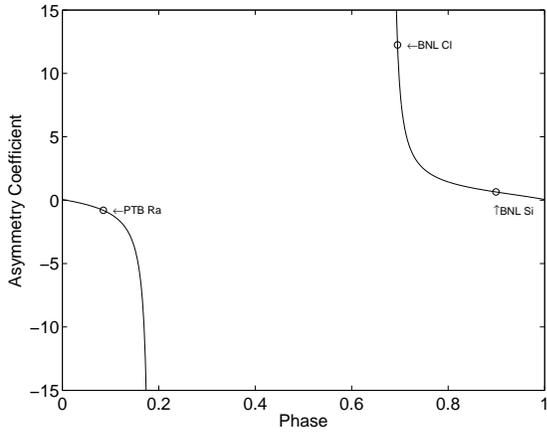}
\caption{Asymmetry coefficient ($A$) as a function of phase ($\phi$), showing the values for BNL $^{36}$Cl, BNL $^{32}$Si, and PTB $^{226}$Ra.\label{fig1}}
\end{figure}

This coefficient is shown as a function of phase in Figure \ref{fig1}. The coefficient is negative for $\kappa=0$ to $\phi_A$ and positive for $\kappa=\phi_A+0.5$ to 1. If $\kappa=\phi_o$, the coefficient is zero, as we would expect. If $\kappa=\phi_A$, the coefficient is infinite, again as we would expect. 

We also find from Eq. \ref{eq6} that the coupling coefficient $\Gamma$ is given by

\begin{equation}
\Gamma=\left(K/C\right)G\left(\kappa\right),
\label{eq8}
\end{equation}

where the ``coupling factor'' $G\left(\kappa\right)$ is given by

\begin{equation}
G=-\frac{\sin\left[2\pi{}\left(\kappa{}-\phi_A\right)\right]}{\sin\left[2\pi{}\left(\phi_A-\phi_o\right)\right]}.
\label{eq9}
\end{equation}

\begin{figure}[ht]
\plotone{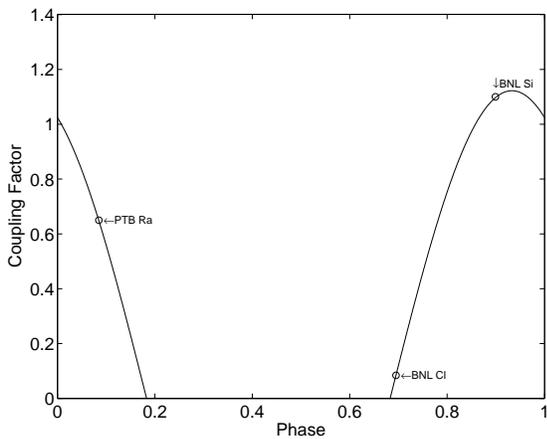}
\caption{Coupling factor ($G$) as a function of phase ($\phi$), showing the values for BNL $^{36}$Cl, BNL $^{32}$Si, and PTB $^{226}$Ra.\label{fig2}}
\end{figure}

This factor is shown as a function of phase in Figure \ref{fig2}. If $\kappa=\phi_o$, the factor is unity. If $\kappa=\phi_A$ or $\phi_A+0.5$, the factor is zero. 

Figures \ref{fig1} and \ref{fig2} confirm that there are no valid solutions to the relevant equations for $\kappa$ in the range $\phi_A<\kappa<\phi_A+0.5$.




\section{Data Analysis}\label{sec:data}

We now compare this model with BNL and PTB data. In our earlier power spectrum analysis \citep{stu10a}, we followed \citet{alb86} and considered only the ratio of the $^{36}$Cl and $^{32}$Si decay rates. However, since we are now interested primarily in the phases of the decay-rate oscillations (rather than their amplitudes), it is necessary to consider the two datasets separately.

\begin{figure}[ht]
\plotone{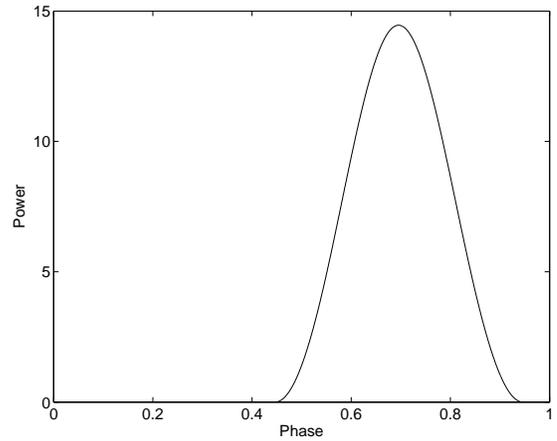}
\caption{Power ($S$) as a function of phase ($\phi$) for BNL $^{36}$Cl data, for frequency 1 yr$^{-1}$.\label{fig3}}
\end{figure}

\begin{figure}[ht]
\plotone{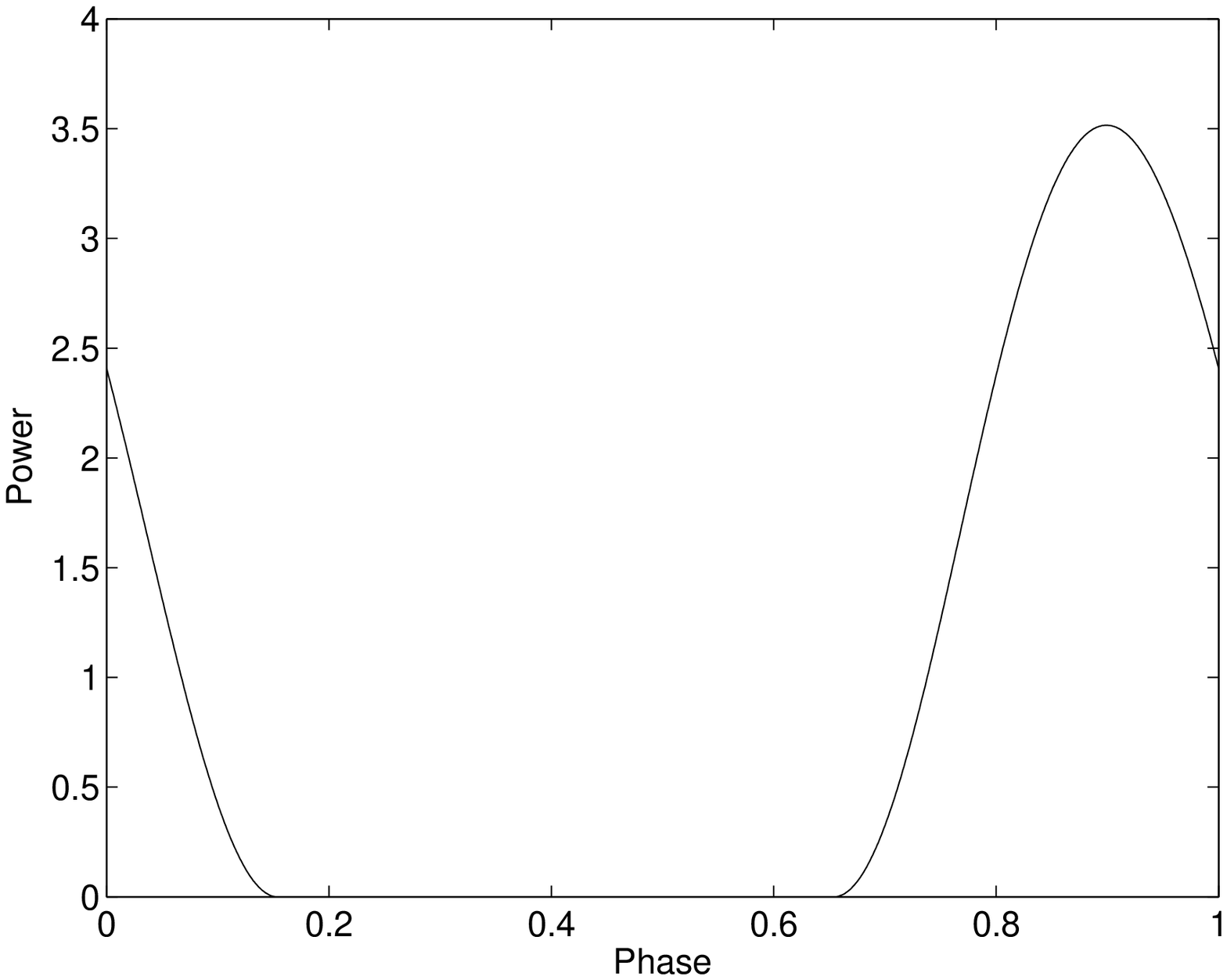}
\caption{Power ($S$) as a function of phase ($\phi$) for BNL $^{32}$Si data, for frequency 1 yr$^{-1}$.\label{fig4}}
\end{figure}

\begin{figure}[ht]
\plotone{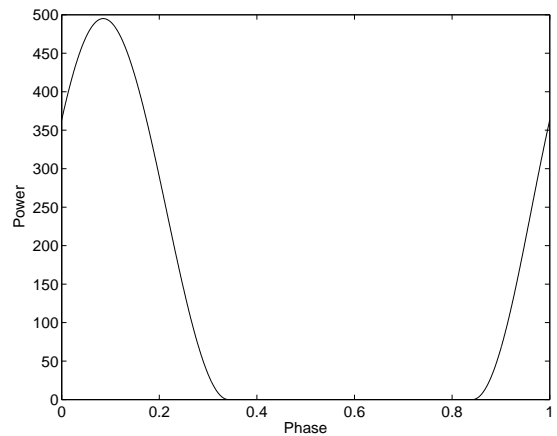}
\caption{Power ($S$) as a function of phase ($\phi$) for PTB $^{226}$Ra data, for frequency 1 yr$^{-1}$.\label{fig5}}
\end{figure}

We determine the phase of the maximum in the annual variation of each dataset by using a likelihood method previously introduced for the analysis of solar neutrino data \citep{stu05}. However, rather than scan the power $S$ as a function of frequency, we now set the frequency at 1 yr$^{-1}$ and scan the power as a function of phase $\phi$. We show the results in Figures \ref{fig3}, \ref{fig4}, and \ref{fig5} for BNL $^{36}$Cl, BNL $^{32}$Si, and PTB $^{226}$Ra, respectively. The results are summarized in Table \ref{tbl:1} where we list, for each element of each experiment, the phase $\phi_P$ of the maximum power, the maximum power $S_P$, the amplitude of the modulation, and the phases $\phi_L,\phi_U$,   for which the power is less than the maximum by 0.5. The phases  and $\phi_L,\phi_U$ denote the ``1-sigma'' offsets.

\begin{table*}
\begin{center}
\caption{Phase, power and amplitude estimates of annual decay-rate variations in the BNL and PTB experiments.\label{tbl:1}}
\begin{tabular}{crrrrrr}
\tableline\tableline
Experiment & Element & $\phi_P$ & $S_P$ & Amplitude & $\phi_L$ & $\phi_U$ \\
\tableline
BNL & $^{36}$Cl & 0.695 & 14.46 & 5.29$\times{}10^{-4}$ & 0.688 & 0.722 \\
BNL & $^{32}$Si & 0.899 & 3.52 & 2.72$\times{}10^{-4}$ & 0.836 & 0.962 \\
PTB & $^{226}$Ra & 0.085 & 495.05 & 8.40$\times{}10^{-4}$ & 0.080 & 0.090 \\
\tableline
\end{tabular}
\end{center}
\end{table*}


	For $^{32}$Si and $^{226}$Ra, we can state with some confidence that the phases fall in the allowed range. However, we see from Figure \ref{fig3} that, for $^{36}$Cl, we can state only that there appears to be 50\% probability that the phase is in the allowed range. An interesting fact to note is that the $^{32}$Si and $^{36}$Cl were measured concurrently, in an alternating fashion, on the same instrument. The fact that these two nuclides do not show the same phase properties further supports the consideration that the observed effects are not systematic in nature, as well as reinforcing the notion that different isotopes will have different inherent sensitivities.

\section{Discussion} \label{sec:disc}


It is important to note that our model refers implicitly to the timing of the unknown solar radiation responsible for the decay-rate variations. Since there may be a time lag between the arrival of the radiation and the emission of a beta particle, there may be a discrepancy between the timings of the two events. This point is relevant to the BNL measurements of the $^{36}$Cl and $^{32}$Si count rates summarized in Table \ref{tbl:1}.

What is experimentally observed in $^{32}$Si decay is the 1709 keV beta particle emitted by $^{32}$P in the sequence $^{32}\rm{Si}\rightarrow^{32}\rm{P}\rightarrow^{32}\rm{S}$, since the 225 keV beta particle emitted from the   transition has too low an energy to have been detected by the BNL experiment. Since the half-life of $^{32}$P is 14.28 days, $^{32}$P comes into equilibrium with the parent $^{32}$Si on a time scale of order $14.28\rm{d}/\rm{ln}\left(2\right)$. Hence the response of the beta-detector lags the initial $^{32}$Si decay by about 21 days. By contrast, $^{36}$Cl decays directly to $^{36}$Ar with the emission of a 709 keV beta particle. Hence the calculated phase difference between the $^{32}$Si and $^{36}$Cl measurements of about 75 days, shown in Table \ref{tbl:1}, is probably due in part to the mechanisms responsible for the decay-rate variations. 

More generally, we draw attention to the following points: The phase distribution that follows from our model rests on the assumptions that decay-rate variations are due entirely to an unknown solar radiation, and that the axis of rotation of the core is parallel to that of the photosphere. If we find that, for some isotopes, the timing of decays does not conform to the expectations of this model, we must question these assumptions. One would then need to consider the possibilities that decays may be due in part to a cosmological particle or radiation flux, as has been suggested by \citet{bau07}, or that the axis of rotation of the solar core is not parallel to that of the photosphere.

As we see from Table \ref{tbl:1}, there is a huge variation in maximum power among the three cases, but a comparatively small range in amplitude. The very large power for the PTB case is due in part to the longer duration of the experiment (14.8 years for PTB as against 7.8 years for BNL). However, even allowing for this difference, we see that different elements have very different sensitivities to whatever is causing the annual modulation. As noted above, this is consistent with expectations from standard weak interaction theory, as noted by \citet{jen10}. 

For the purposes of this article, we refer to ``decays'' as if a decay event is a simple process, but this is typically not the case. For instance, the decay of $^{226}$Ra is a complex process. Although the decay of a $^{226}$Ra nucleus begins with alpha emission, and although after about 20 years $^{226}$Ra nuclei are in secular equilibrium with the majority of its 13 radioactive daughters, it takes about 200 years for the processes to effectively reach equilibrium \citep{chr83,chi07}. When equilibrium is reached, ~42\% of the photon emissions are due to beta-decaying daughters. The ionization chamber utilized in the PTB experiment cannot discriminate between either (alpha or beta) type of decay: the chamber measures only the total energy deposited by the incident photons, which have their origins in both types of decay from several different isotopes.

The asymmetry coefficient and the coupling factor, given by Eqs. \ref{eq7} and \ref{eq9}, respectively, are listed in Table \ref{tbl:1} and shown in Figures \ref{fig1} and \ref{fig2}, respectively. We see that the phases fall in the ``allowed'' range 0.683 to 0.183 (avoiding the ``forbidden'' range 0.183 to 0.683). We identify in these figures the phases of maximum power listed in Table \ref{tbl:1}. On applying the results shown in these two figures to the data shown in Table \ref{tbl:1}, we obtain the values of the asymmetry coefficient and of the coupling coefficient listed in Table \ref{tbl:2}.

\begin{table*}
\begin{center}
\caption{Estimates of the asymmetry coefficients ($A$) and coupling coefficients ($\Gamma$) derived \\ from data for the BNL and PTB experiments.\label{tbl:2}}
\begin{tabular}{cccccc}
\tableline\tableline
 &  &  &  & Asymmetry & Coupling  \\
Experiment & Element & Amplitude & Phase & Coefficient & Coefficient \\
\tableline
   &  &  & 0.688 & N/S & N/S \\
BNL & $^{36}$Cl & 5.29$\times{}10^{-4}$ & 0.695 & 12.25 & 0.0013  \\
   &  &  & 0.722 & 4.02 & 0.0043 \\
   &  &  &  &  &  \\
   &  &  & 0.836 & 1.08 & 0.0076 \\
BNL & $^{32}$Si & 2.72$\times{}10^{-4}$ & 0.899 & 0.65 & 0.0090  \\
   &  &  & 0.962 & 0.29 & 0.0090 \\
   &  &  &  &  &  \\
   &  &  & 0.080 & -0.72 & 0.0171 \\
PTB & $^{226}$Ra & 8.40$\times{}10^{-4}$ & 0.085 & -0.80 & 0.0163  \\
   &  &  & 0.090 & -0.89 & 0.0156 \\
\tableline
\end{tabular}
\end{center}
\end{table*}

In reviewing the results shown in Table \ref{tbl:2}, we see that the results for the $^{36}$Cl experiment are ill-determined, since the phase is close to the critical phase 0.683. In comparing the results for the BNL $^{32}$Si experiment and the PTB $^{226}$Ra experiment, we see that there is no substantial difference between the estimates of the magnitudes of the asymmetry and coupling coefficients, despite the fact that the powers differ greatly. It is interesting to note the difference in sign between these two experiments: for both elements investigated in the BNL data, the modulation shows evidence of a larger influence arising from the northern solar hemisphere, whereas for the PTB experiment, the modulation shows evidence of a larger influence arising from the southern hemisphere. Since the time intervals of the two experiments are different (1982.13 to 1989.93 for BNL and 1983.86 to 1999.17 for PTB), we cannot infer whether the difference is due to variability of the solar influence, or to a difference in the ``sensitivity'' of the specific decaying isotopes. It will therefore be important to study a variety of isotopes concurrently for precisely the same time interval. A framework to describe a possible class of mechanisms is presented in \citet{fis09}.

We have found evidence, reported in previous publications \citep{stu10a,stu10b}, of modulations in decay rates that may be attributable to rotation at or near the solar core. We have also recently found evidence of a Rieger-type modulation that we may attribute to an ``inner tachocline'' separating the core from the radiative zone \citep{stu11}. These results show that whatever radiation is responsible for influencing decay rates propagates essentially freely through the solar radiative and convection zones, suggesting that some flavor of neutrinos may be responsible for decay-rate variations. However, the standard theory of neutrino physics does not provide for a mechanism by which neutrinos could have such an effect, raising the possibility of new physics.

In view of these conflicting considerations, it is clearly desirable that we obtain further information relevant to the mechanism by which the Sun influences decay rates. If neutrinos or some other particles (or quanta of radiation) influence radioactive atoms, it is conceivable that they may transfer their momentum to the decaying nuclei, and thus exert a force on a bulk sample of radioactive material. This raises the possibility of carrying out a ``fifth force'' type of experiment on a macroscopic radioactive sample, using techniques recently developed to search for deviations from Newtonian gravity \citep{fis99}.

\acknowledgments

We are indebted to D. Alburger and G. Harbottle for supplying us with the BNL raw data, and to H. Schrader for supplying us with the PTB raw data. We are also indebted to an anonymous referee for suggestions that materially improved the article. The work of PAS was supported in part by the NSF through Grant AST-06072572, and that of EF was supported in part by U.S. DOE contract No. DE-AC02-76ER071428.

\end{document}